\documentclass[12pt,preprint]{aastex}
\usepackage{amssymb}

\newcommand{\xmm}{{\it XMM-Newton}}

\newcommand{\dt}{\Delta t}
\newcommand{\numin}{\nu_{\rm min}}
\newcommand{\numax}{\nu_{\rm max}}
\newcommand{\fb}{\nu_{\rm b}}
\newcommand{\lbol}{L_{\rm bol}}

\newcommand{\psdamp}{{\rm {PSD_{amp}}}}
\newcommand{\tb}{T_{\mathrm{b}}}

\newcommand{\lfb}{\nu_{\rm LFB}}
\newcommand{\cb}{C_{\mathrm{b}}}
\newcommand{\clfb}{C_{\mathrm{LFB}}}
\newcommand{\mbh}{M_{\rm BH}}
\newcommand{\msun}{M_{\odot}}
\newcommand{\me}{\dot{\rm m}_{\rm E}} 
\newcommand{\nxs}{\sigma^2_{\rm NXS}}

\newcommand{\nxsb}{\sigma^2_{\rm NXS,b}}

\newcommand{\nxsd}{\sigma^2_{\rm NXS,300d}}

\newcommand{\et}{et al.\ }

\shortauthors{Zhang}

\shorttitle{X-ray variability and accretion state of AGN}

\begin{document}

\title{The longest timescale X-ray variability reveals an evidence for active galactic nuclei in the high accretion state}

\author{You-Hong Zhang}
\affil{Department of Physics and Tsinghua Center for Astrophysics
(THCA), Tsinghua University, Beijing 100084, China}
\email{youhong.zhang@mail.tsinghua.edu.cn}

\begin{abstract}
The All Sky Monitor (ASM) onboard the Rossi X-ray Timing Explorer
(RXTE) has continuously monitored a number of Active galactic nuclei
(AGNs) with similar sampling rates for 14 years from 1996 January to
2009 December. Utilizing the archival ASM data of 27 AGNs, we
calculate the normalized excess variances of the 300-day binned
X-ray light curves on the longest timescale (between 300~days and
$14$~years) explored so far. The observed variance appears to be
independent of AGN black hole mass and bolometric luminosity,
respectively. According to the scaling relation with black hole mass
(and bolometric luminosity) from Galactic black hole X-ray binaries
(GBHs) to AGNs, the break timescales which correspond to the break
frequencies detected in the power spectral density (PSD) of our AGNs
are larger than binsize (300~days) of the ASM light curves. As a
result, the singly-broken power-law (soft-state) PSD predicts the
variance to be independent of mass and luminosity, respectively.
Nevertheless, the doubly-broken power-law (hard-state) PSD predicts,
with the widely accepted ratio of the two break frequencies, that
the variance increases with increasing mass and decreases with
increasing luminosity, respectively. Therefore, the independence of
the observed variance on mass and luminosity suggests that AGNs
should have the soft-state PSDs. If taking into account the scaling
of breaking timescale with mass and luminosity synchronously, the
observed variances are also more consistent with the soft-state than
the hard-state PSD predictions. With the averaged variance of AGNs
and the soft-state PSD assumption, we obtain a universal PSD
amplitude of $0.030\pm0.022$. By analogy with the GBH PSDs in the
high/soft state, the longest timescale variability supports the
standpoint that AGNs are scaled-up GBHs in the high accretion state,
as already implied by the direct PSD analysis.
\end{abstract}



\section{Introduction}\label{sec:intro}

Active galactic nuclei (AGNs) might be scaled-up Galactic black hole
X-ray binaries (GBHs) (e.g., McHardy et al. 2006) since both are
powered by similar physical process, i.e., accretion onto
(super-massive and stellar-mass, respectively) black hole.
Therefore, it is important and fundamental to explore the
observational similarities between the two classes of black hole
accreting system on very different scales. Because it originates in
the region mostly close to the black hole, the X-ray emission
carries the most pivotal messages for the black hole accretion
process. Both AGNs and GBHs exhibit strong X-ray variability, which
is usually quantified by the power spectral density (PSD;
$P_{\nu}\propto\nu^{\alpha}$) of their X-ray light curves.

The GBH PSDs between 0.001 and 100~Hz are closely related to their
X-ray spectral states (e.g., Done \& Gierli\'nski 2005; Klein-Wolt
\& van der Klis 2008). A template source mostly quoted is the
persistent source, Cyg~X-1. In its low/hard state at which the
energy spectrum is dominated by a strongly variable power-law
component, the hard-state PSD of Cyg~X-1 is approximately described
by a doubly-broken power law (e.g., Pottschmidt et al. 2003),
flattening from $\alpha\sim -2$ to $\alpha\sim -1$ at the
(high-)frequency break ($\fb \sim$ a few Hz) and further flattening
to $\alpha\sim 0$ at the low-frequency break ($\lfb \sim$ a few
tenth Hz) from high to low frequencies. The ratio of the two break
frequencies is $\sim10-100$. In its high/soft state, albeit the
energy spectrum is dominated by a roughly constant thermal disc
component, the highly variable power-law component is strong enough
to construct the soft-state PSD of Cyg~X-1, which is significantly
distinguished from the hard-state PSD by showing only one
(high-frequency) break ($\fb \sim15$~Hz). The soft-state PSD slope
above and below $\fb$ is $\alpha\sim -2$ and $\alpha\sim -1$,
respectively, and the $-1$ slope can be extended down to many
decades of frequencies without further flattening (e.g., Axelsson,
Borgonovo, \& Larsson 2006). However, as already shown by
Pottschmidt et al. (2003), even in Cyg~X-1, the GBH PSD shape is not
as simple as usually assumed.

If the X-ray variability timescale linearly scales with black hole
mass ($\mbh$) from GBHs to AGNs, the state-transition timescales of
a few days found in GBHs ($\mbh\sim10\msun$) suggest that AGNs of
$\mbh\sim10^7\msun$ would show state transitions on timescales of a
few thousand years. Thus, the state transitions of AGNs can not be
verified by direct X-ray observations if AGNs do have state
transitions behaving like GBHs and satisfy the generally accepted
scaling law. Nevertheless, if we believe that AGNs have the same
underlying variability mechanism as GBHs, the different sub-classes
of AGNs might be reasonably assumed to correspond to different
spectral states of GBHs. In fact, this assumption can be tested by
comparing the PSDs of different types of AGNs with those of GBHs in
different spectral states. However, the scaling of timescale with
mass ascertain that it is very time-consuming to derive AGN PSDs.
Compared to GBHs, the limited durations and low signal-to-noise
ratios of AGN light curves usually produce incomplete and low
quality PSDs. As a result, it is not easy to constrain PSD shapes
and the important break frequencies for a large number of AGNs.
Relatively high quality AGN PSDs obtained so far, actually available
for a limited number (about a dozen) of Seyfert galaxies only, are
almost analogous to GBH soft-state PSDs (e.g., Uttley \& McHardy
2005). The PSDs of Seyfert galaxies show one break ($\fb$) only,
flattening from $\alpha \sim -2$ to $\alpha \sim -1$. With present
data, the PSDs do not show further flattening down to the lowest
frequencies ($\sim 10^{-8}$~Hz) which are $3-4$ decades lower than
the break frequencies. Like GBHs in the high/soft state, the fact
that Seyfert galaxies show the soft-state PSDs are strongly
supported by their low radio fluxes and high accretion rates larger
than two percent of the Edington accretion rate at which Cyg~X-1
transits between the hard and soft spectral states (e.g., Maccarone
et al. 2003).

Up to now, Ark~564 and NGC~3783 are the only AGNs whose PSD shows
the second, low-frequency break ($\lfb$). Ark~564 has strong
evidence for the second break in its PSD (Pounds et al. 2001;
Papadakis \et 2002; Markowitz et al. 2003; McHardy et al. 2007).
Nevertheless, the very high (possibly super-Eddington) accretion
rate ($\me \sim 1$, Romano et al. 2004) suggests that Ark~564 may
resemble GBHs in the very high state rather than in the low/hard
state (McHardy et al. 2007), since in the very high state GBH PSDs
show two distinct breaks as well. The second break in the PSD of
NGC~3783, firstly presented by Markowitz et al. (2003), was
questioned by Summons et al. (2007) with the improved PSD. Moreover,
the moderately high accretion rate ($\me \sim 0.07$, Uttley \&
McHardy 2005) and radio quietness (e.g., Reynolds 1997) suggest that
NGC~3783 is analogous to GBHs in the high/soft state. Therefore,
NGC~3783 also became one of AGNs showing soft-state PSD, which
leaves Ark~564 as the only non-soft-state AGN.

The X-ray variability of AGNs can also be quantified with the
normalized excess variance ($\nxs$) of their X-ray light curves,
which is approximately equal to the integral of the normalized PSDs
of the same light curves (e.g., Vaughan et al. 2003; Zhang \et
2005). Because it is much more easier to calculate the variances
than to calculate the PSDs, the variances have been used to explore
the scaling relation of the X-ray variability with black hole mass,
i.e., the so-called variance-mass ($\nxs-\mbh$) relationship, and
X-ray luminosity of AGNs. On timescale of about half day, the
variance anti-correlates with mass (i.e., $\nxs \propto \mbh ^{-1}$;
e.g., Lu \& Yu 2001; Bian \& Zhao 2003; Liu \& Zhang 2008; Zhou et
al. 2010) and X-ray luminosity (e.g., Nandra et al. 1997). By
scaling with Cyg~X-1, the variance has been used as a method to
estimate $\mbh$ of AGNs (Nikolajuk et al. 2004; 2006; 2009;
Gierli\'nski et al. 2008; Zhang et al. 2005). Although the variance
of an individual source does nothing for constraining AGN PSDs, the
variance-mass relation, just requiring to calculate variances of a
large sample of AGNs with known masses, has this function. If AGNs
have the same underlying X-ray variability mechanism that scales
with mass, the variance-mass relation can be reproduced from a
universal AGN PSD model. The comparison between the observed and
predicted variance-mass relation can determine AGN PSD shape and the
scaling factors of its parameters with mass. For 10 Seyfert
galaxies, which have been regularly observed in X-rays, the
2--10~keV variances calculated with 300-day long RXTE PCA light
curves anti-correlate with masses, which can be fitted by the
hard-state PSD model (Papadakis 2004). With ASCA light curves of
about half-day long for a larger sample of AGNs, O'Neill \et (2005)
obtained a similar anti-correlation between variances and masses,
which is also in agreement with the prediction of the hard-state PSD
model. Using $\xmm$ data on timescale of about half day, Papadakis
\et (2008b) and Miniutti \et (2009) enlarged the AGN sample,
especially objects with smaller masses, for the variance-mass
relation. Papadakis \et (2008a) also extended the study of the
variance-mass relation to higher red-shift AGNs.

The studies of variance-mass relationship suggest that AGNs have the
hard-state PSDs, which apparently is in contradiction with the
conclusion that AGNs have the soft-state PSDs from the direct PSD
analysis. Importantly, the AGN samples of Papadakis (2004) and
O'Neill \et (2005) include most of AGNs whose soft-state PSDs have
been well determined. The reason for this inconsistency could be
mainly due to the short duration of the light curves used to
calculate the variances. In fact, even though AGNs do have the
hard-state PSDs, the short timescale variances do not cover the
frequencies where the low-frequency breaks locate for most of AGNs.
On short timescales, the hard-state PSD predicts the same
variance-mass relation as the soft-state PSD does. Short timescale
variance-mass relation is thus not able to differentiate the two PSD
models. Consequently, in order to demonstrate whether AGN PSDs have
the second breaks, it is necessary to obtain variance-mass relation
with very long timescale light curves, especially for AGNs with
large masses. The values of $\lfb$ for the GBH hard-state PSDs are
about a few tenth of one Hz. If AGNs do show the hard-state PSDs, by
scaling law of $\lfb$ with mass, $\lfb$ would be $\sim10^{-7}$~Hz
(on the order of one year) for AGNs with $\mbh\sim10^7\msun$.
Therefore, in order to effectively distinguish the two AGN PSD
models with variance-mass relation, a large number of AGNs should be
continuously monitored for tens of years.

The longest data stream with roughly continuous and regular sampling
mode are from the ASM onboard RXTE. The ASM has been monitoring a
large number of AGNs since January 1996. In this paper, we will use
the ASM data to obtain the variances of 27 AGNs on the longest
timescale so far, with which we are in an attempt to differentiate
the two AGN PSD models.


\section{The ASM Data and AGN Sample}\label{sec:obs}

The ASM operates in the 1.5 to 12 keV energy band and scans most of
the sky every 1.5 hours. We start our analysis with the data of the
one-day averages from the ASM quick-look
pages\footnote{http://xte.mit.edu/ASM$\_$lc.html}. Each one-day
averaged data is the average of the fitted source fluxes from a
number (typically 5 to 10) of individual 90 second dwells over that
day, and is quoted as the nominal 2--10~keV ASM count rate (counts
per second). We make use of the full ASM data stream from 1996
January 1 to 2009 December 31 (Modified Julian Day [MJD]
$50083-55196$), which provides 14-year long X-ray light curves for a
large number of AGNs. In the ASM quick-look pages, we select AGNs
having a measurement of black hole mass in the literatures. Our
final sample, tabulated in Table~\ref{tab:agn}, contains 27 AGNs,
including 20 broad-line Seyfert 1 (BLS1) and 7 narrow-line Seyfert 1
(NLS1) galaxies. Among them, we find bolometric luminosity of 20
AGNs and PSD break timescale ($1/\fb$) of 12 AGNs in the
literatures.

\begin{figure} \epsscale{1.0}
\plottwo{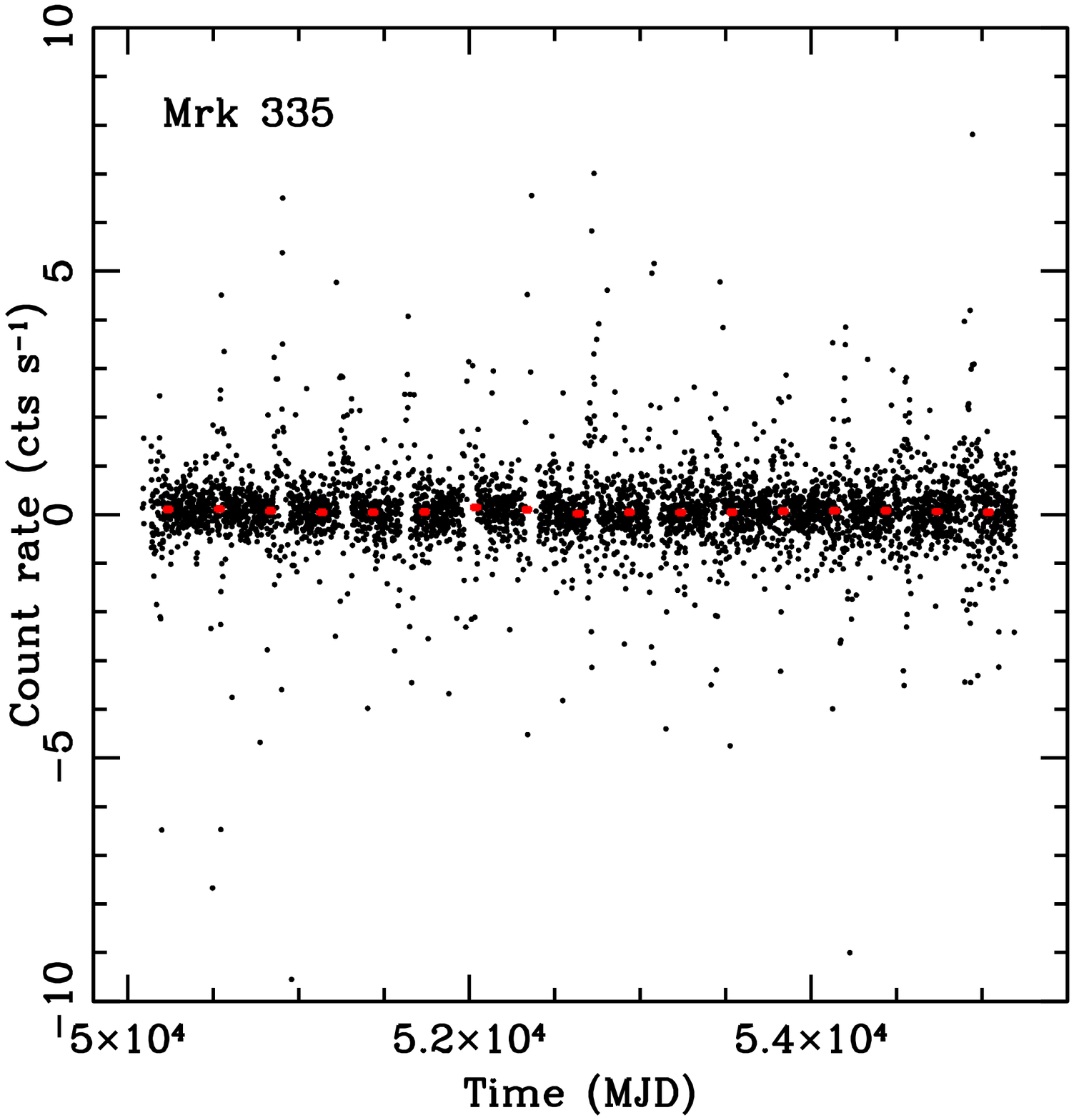}{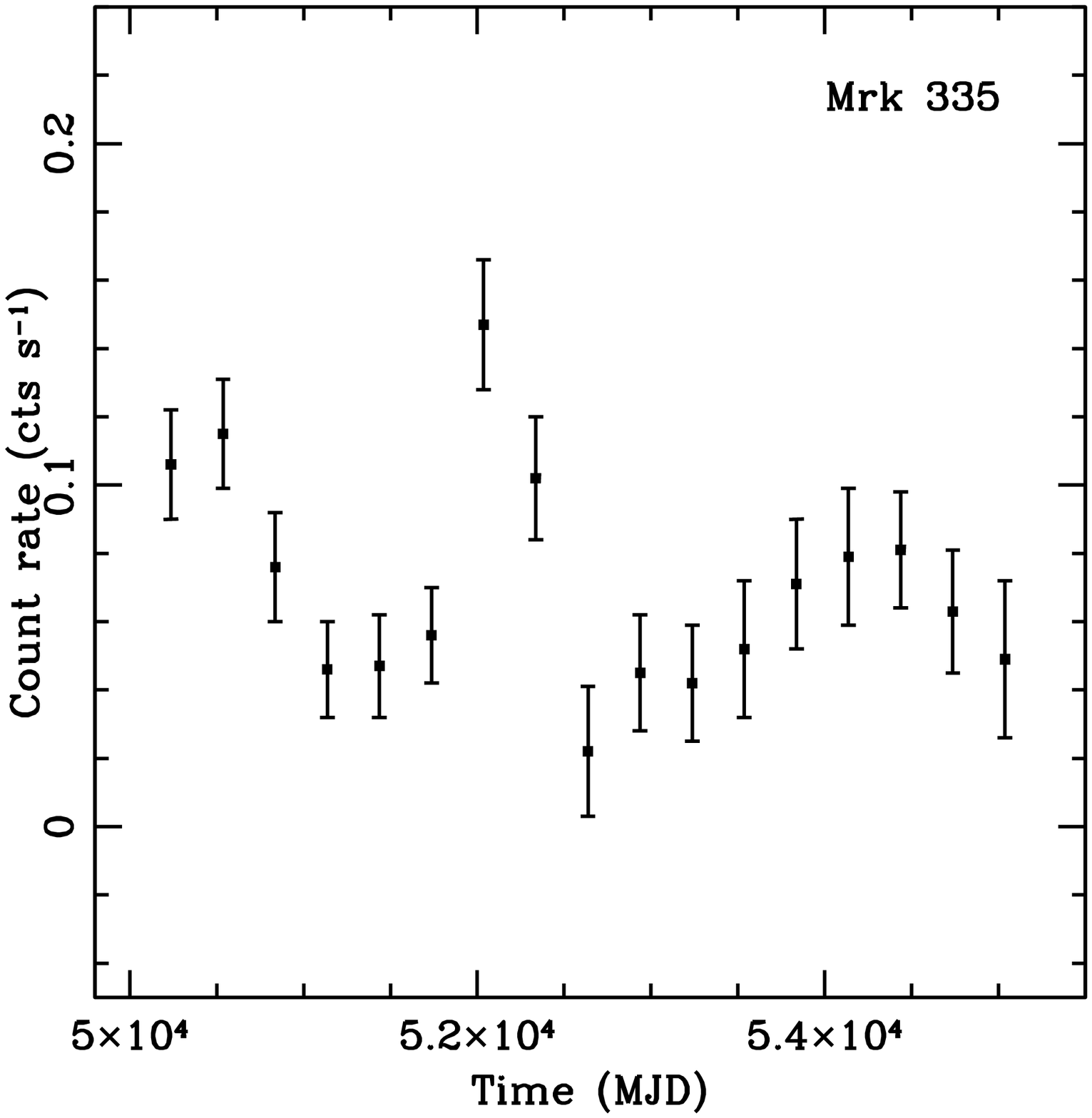} \caption{ \footnotesize The
ASM light curve for the lowest count rate object Mrk~335. The left
plot shows the one-day averaged light curve (black solid circles
whose errors are not shown for clarity) with the 300-day averaged
light curve (red solid squares whose errors are smaller than the
symbol size) plotted on top. The right plot presents again the
300-day averaged light curve for clearly showing the source's
variations on timescale of 300~days.} \label{fig:mrk335}
\end{figure}

\begin{figure} \epsscale{1.0}
\plottwo{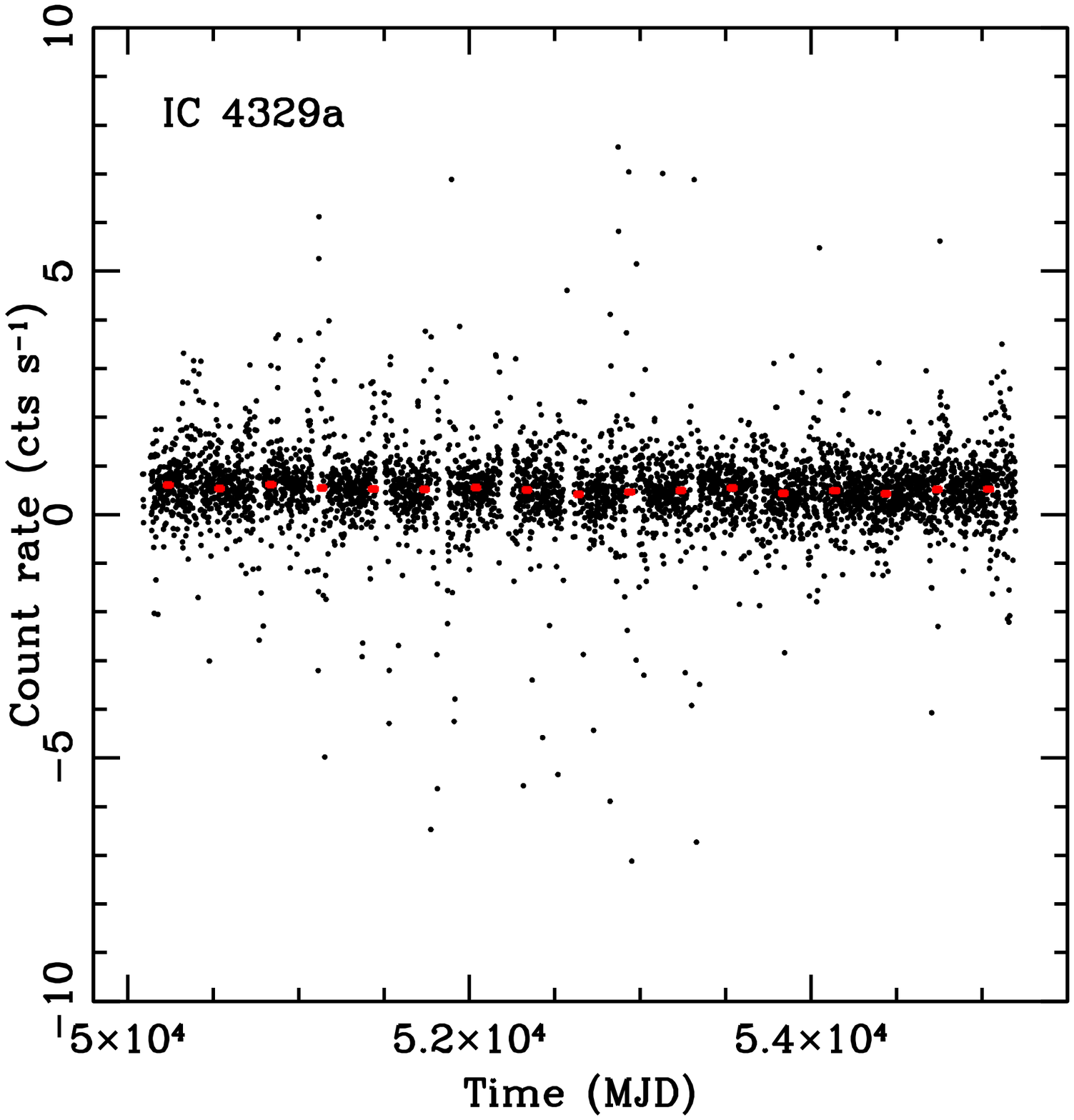}{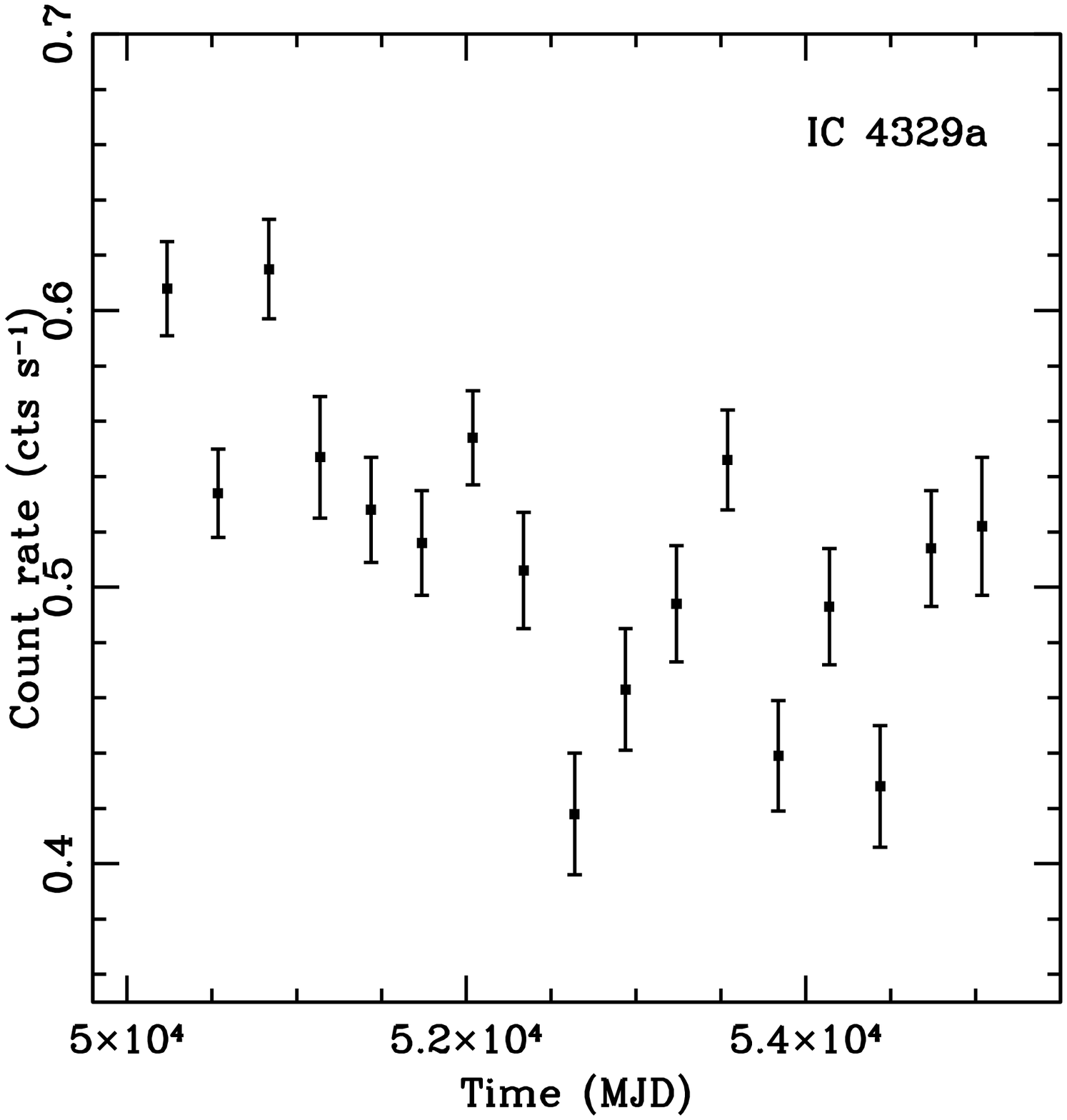} \caption{ \footnotesize
Same to Figure~\ref{fig:mrk335} but for the highest count rate
object IC~4329a. } \label{fig:ic4329a}
\end{figure}

Due to low sensitivity, the ASM signal-noise ratios are quite low
for AGNs. As two examples, the left plots of Figure~\ref{fig:mrk335}
and Figure~\ref{fig:ic4329a} show the background-subtracted one-day
averaged ASM light curves for the lowest and highest count rate
object Mrk~335 and IC~4329a among our AGN sample, respectively. It
can been seen that the one-day averaged light curves contain many
negative count rates. The light curves do not show legible
variability trend either, mainly caused by strong Poisson noise. In
order to acquire meaningful variability information of AGNs from the
ASM data, it is necessary to rebin the one-day averaged light curves
over very long timescale (e.g., 300~days as used in
Section~\ref{sec:results}). The right plots of
Figure~\ref{fig:mrk335} and Figure~\ref{fig:ic4329a} present the
300-day averaged light curves of Mrk~335 and IC~4329a, showing that
the two objects are indeed variable on such long timescale, though
the errors are still large. For comparisons, the 300-day averaged
light curves are also plotted on top of their respective one-day
averaged light curves in the left plots of Figure~\ref{fig:mrk335}
and Figure~\ref{fig:ic4329a}.

\begin{figure} \epsscale{1.0}
\plottwo{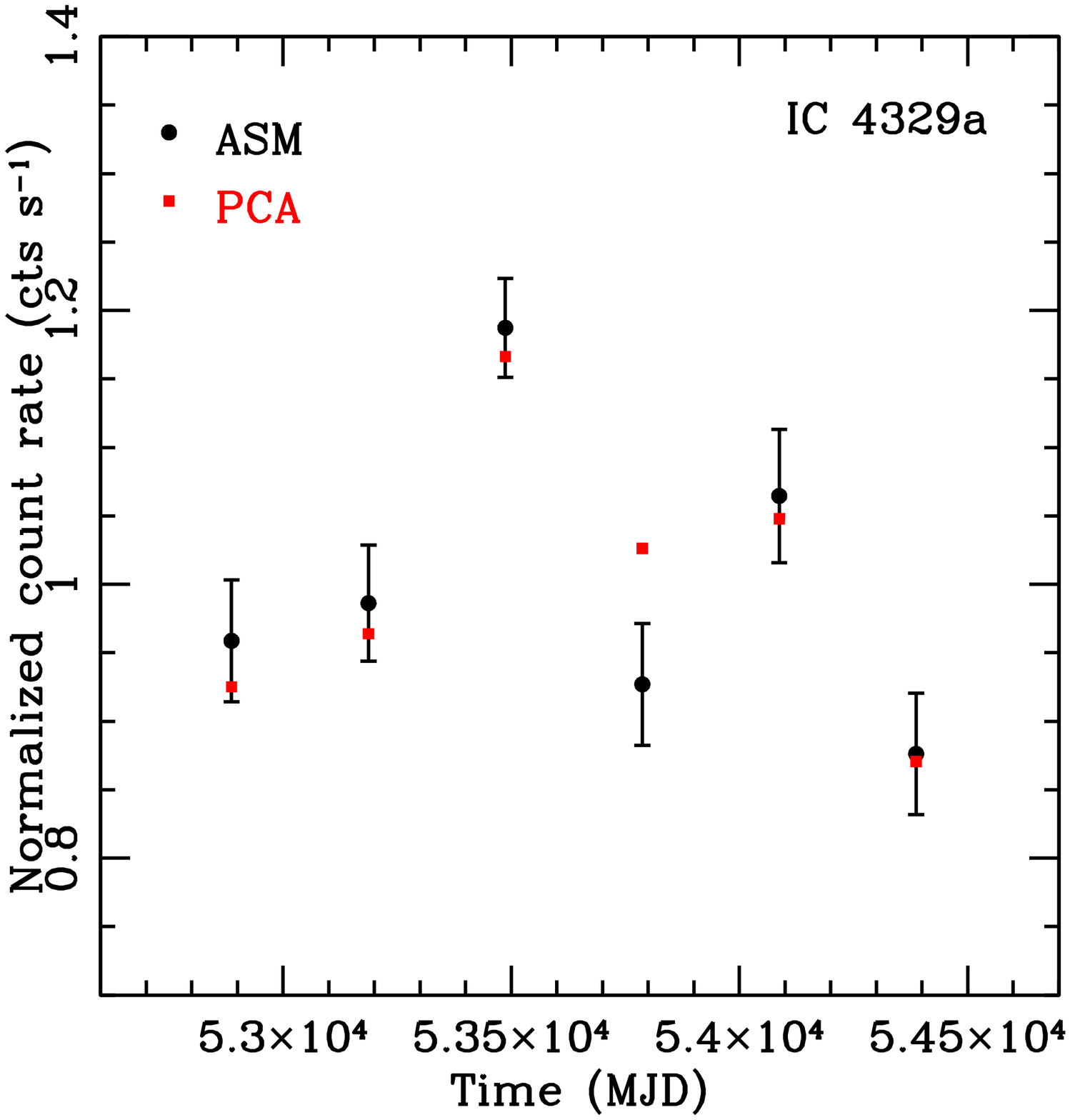}{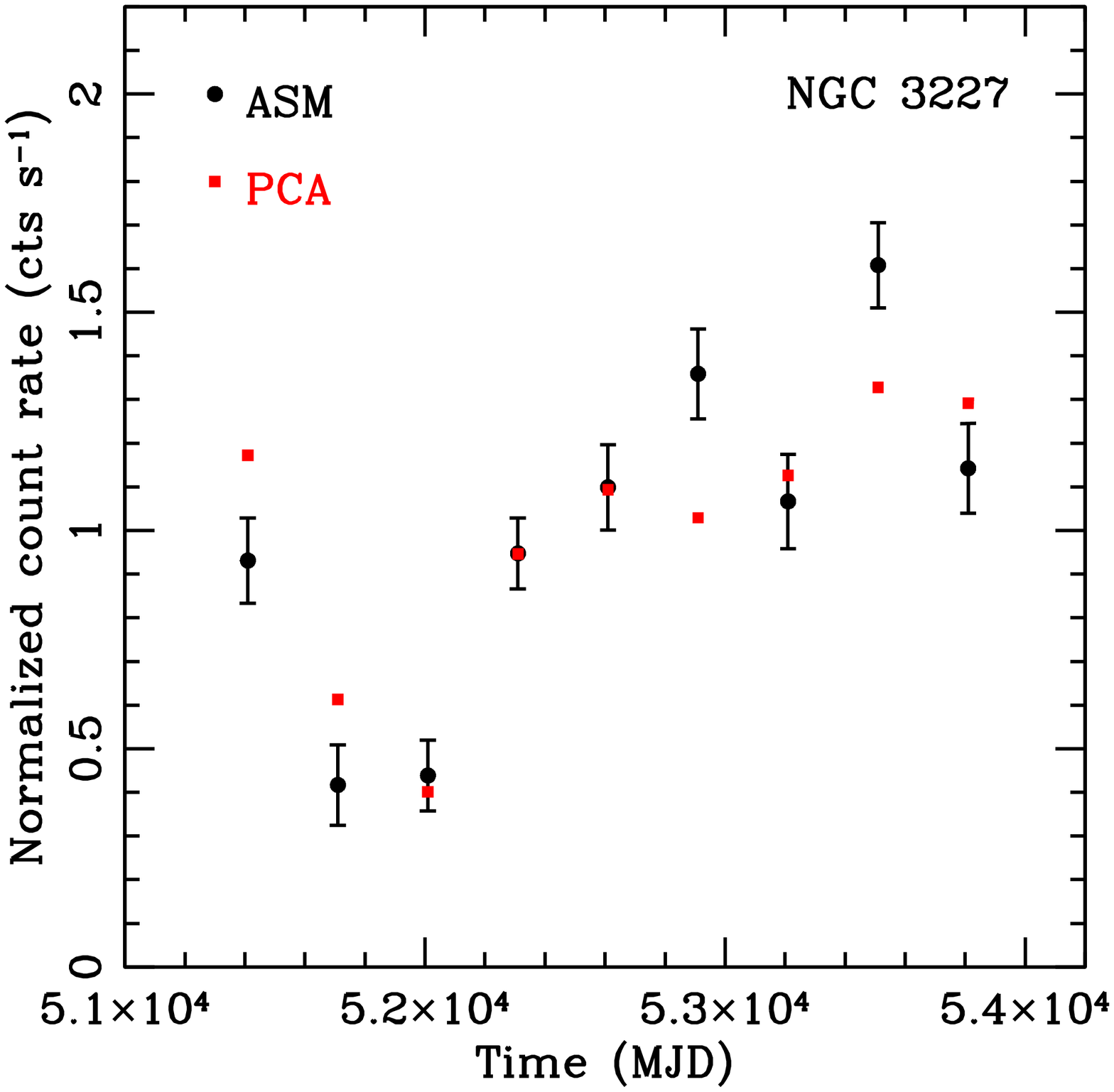} \caption{ \footnotesize The PCA
(red solid squares whose errors are smaller than the symbol size)
and ASM (black solid circles) 300-day averaged light curves in the
time interval over which the long-term PCA observations were
performed. For direct comparisons, both light curves are normalized
to their respective mean count rates. Left plot is for IC~4329a, and
right plot for NGC~3227. Although the differences are present for
some time points, the ASM light curves could be considered to
roughly follow the PCA ones. } \label{fig:pca}
\end{figure}

Onboard RXTE, the Proportional Counter Array (PCA) is much more
sensitive than the ASM. To judge the quality of the ASM light
curves, we compare the light curves obtained with PCA and ASM in the
same time intervals. We choose two objects, namely IC~4329a and
NGC~3227, to perform this comparison. The PCA monitored IC~4329a
once every 4.26 days for a duration of 4.3 years, from 2003 April 8
to 2007 August 7 (MJD $52737-54319$; observation identifiers
80152-03, 80152-04, 90154-01, 91138-01, and 92108-01; see Markowitz
2009). NGC~3227 was monitored by the PCA with different observing
schemes (see Uttley \& McHardy 2005 for the details) for a duration
of 7.4 years, from 1999 March 23 to 2006 August 13 (MJD
$51260-53960$; observation identifiers 40151-01, 40151-08, 40151-09,
50153-07, 50153-08, 60133-05, 70142-01, 80154-01, 90160-04). We
obtained the long-term PCA light curves of the two objects from
HEASARC archive data search
form\footnote{http://heasarc.gsfc.nasa.gov/cgi-bin/W3Browse/w3table.pl},
in which we selected "XTE Target Index Catalog" and downloaded the
merged light curves. We then re-selected the one-day averaged ASM
light curves of the two objects in the same time intervals as
spanned by their long term PCA light curves. Both the PCA and ASM
light curves are averaged over 300~days, which are shown in
Figure~\ref{fig:pca}. In order to compare the PCA and ASM light
curves easily, they are normalized to their respective mean count
rates. Although the discrepancies of the normalized count rates
between the ASM and PCA light curves are present in some points of
the light curves and the PCA errors are much smaller than the ASM
ones, the ASM light curves could still be thought to roughly track
the PCA ones on binsize of 300~days. Therefore, the 300-day averaged
ASM light curves can be used to study long-term variability of AGNs.

\begin{figure} \epsscale{1.0}
\plottwo{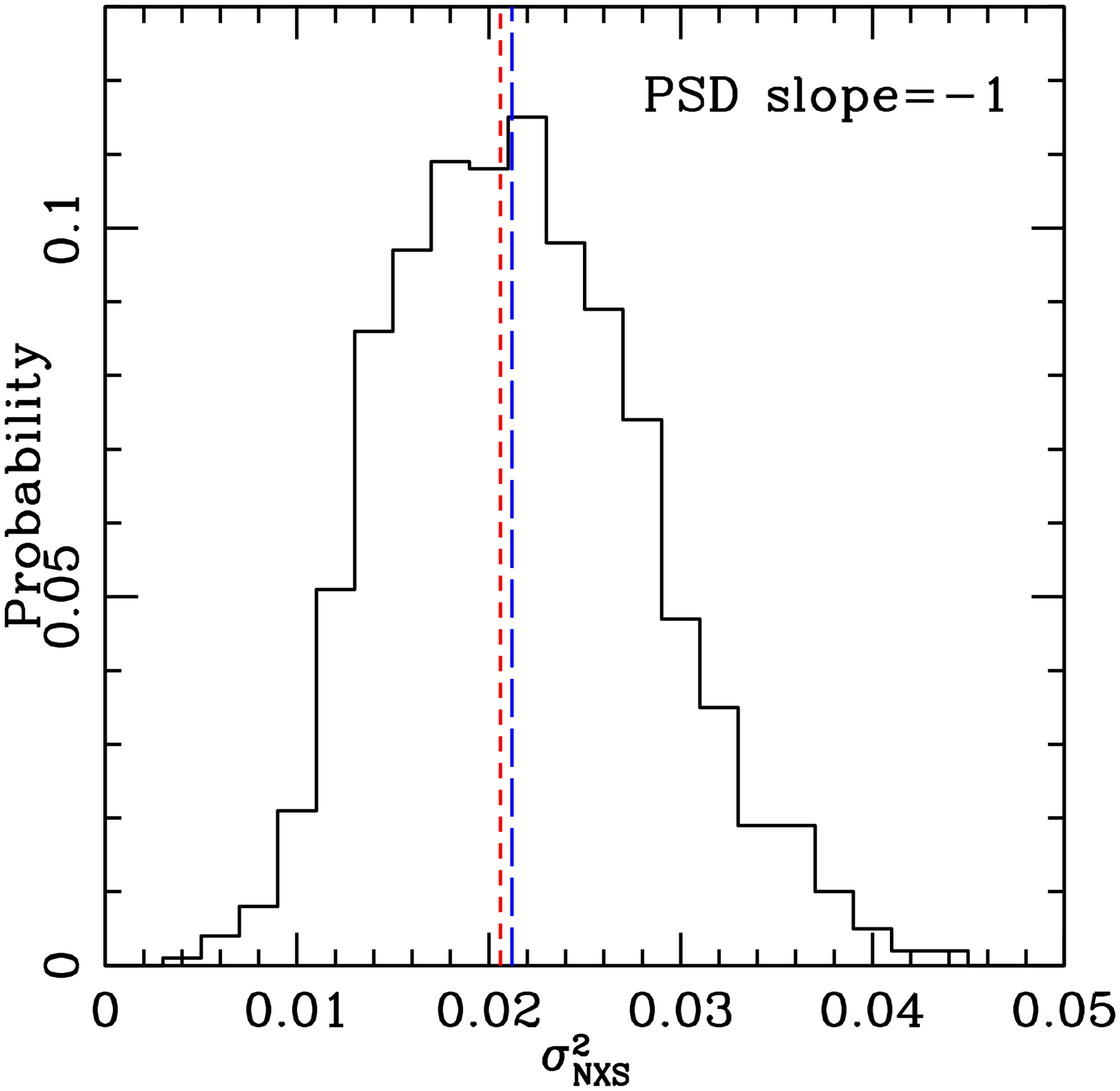}{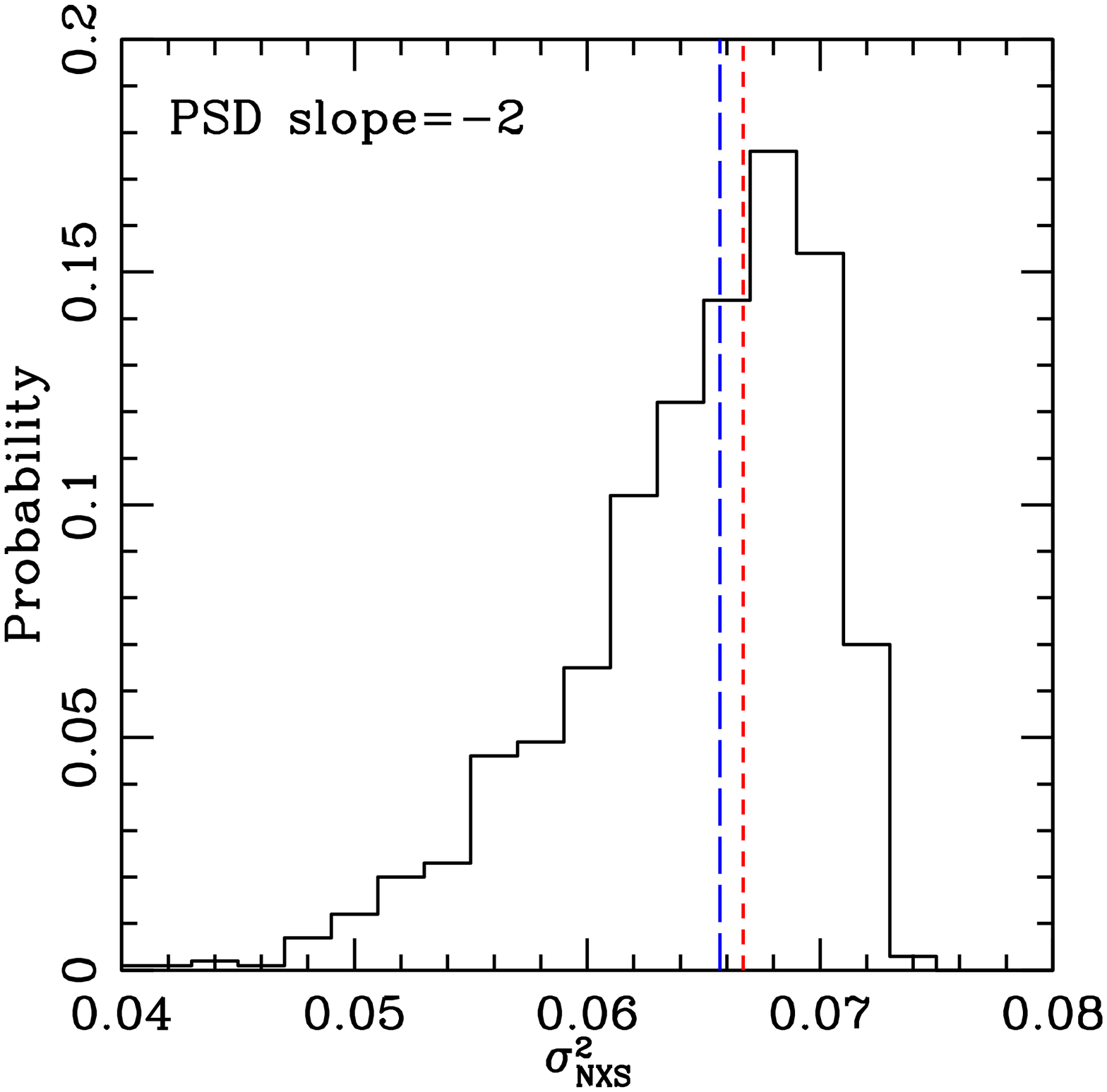} \caption{ \footnotesize The
probability distribution (the black solid line) of $\nxsb$ estimated
from the simulated light curves of 5100-day long and 300-day
binsize. The blue long-dashed line indicates the median of the
distribution. The red short-dashed line presents the value of the
integral of the intrinsic PSD from $\numin = 1/T$ to $\numax =
1/(2\dt)$, where $T=5100$~days and $\dt=300$~days. }
\label{fig:prob}
\end{figure}

In the case of the heavily binned ASM light curves, the sampled PSD
at $\numax$ ($\numax=1/(2\dt)$, where $\dt$ is the binsize of the
light curves) might be massively affected by aliasing effects,
specially because of $\fb \gg \numin$ ($\numin=1/T$, where $T$ is
the duration of the light curves) for most of objects. Therefore,
the observed $\nxs$ (see formula~[\ref{eq:nxs}] for the definition)
is not equal to the integral of the intrinsic PSD between $\numin$
and $\numax$. This should not be a serious problem, as it should
affect all sources in a similar way. Nonetheless, it is worth
investigating this aliasing effect by simulating a red-noise light
curve. We follow the procedure described by Timmer \& K$\ddot{\rm
o}$nig (1995) to fake a light curve ( see Zhang 2002 and Zhang et
al. 2004 for the details) by assuming a single power-law PSD
($P_{\nu} = a \times \nu^{\alpha}$, where $a$ is the PSD
normalization factor). The experiment is performed twice, one with
slope of $\alpha=-1$ and in the second case with $\alpha=-2$. In
order to closely imitate the ASM light curves, we produce a light
curve with 5100 points (one per day). For simplicity, we do not
consider the effect of Poisson noise. The light curve is normalized
to have a mean count rate of 0.2 and a variance of
$7.56\times10^{-2}$. The normalized variance is the integral of the
intrinsic PSD between $1/T$ and $1/(2\dt)$, where $T=5100$~days and
$\dt=1$~day correspond to the length and binsize of the simulated
light curve. Under these assumptions, one can derive the value of
$a$, which is $9.64 \times 10^{-3}$ and $1.72 \times 10^{-10}$ for
$\alpha=-1$ and $\alpha=-2$, respectively. With the known $a$, we
can calculate the value of the integral of the intrinsic PSD between
any frequency range. For the case of $T=5100$~days and
$\dt=300$~days, the integral is $\nxsd=2.06\times10^{-2}$ and
$\nxsd=6.67\times10^{-2}$ for $\alpha=-1$ and $\alpha=-2$,
respectively. The simulated light curve is then rebinned with
$\dt=300$~days (i.e., the binsize of our ASM light curves),
resulting in 17 bins (one per 300~days), and its normalized
variance, $\nxsb$, is estimated. The aliasing effect can be examined
by comparing the value of $\nxsb$ with that of $\nxsd$. The
procedure is repeated by 1000 times, from which we obtain the median
and error ($90\%$ confidence level) of $\nxsb$, which is
$(2.12^{+1.25}_{-0.94}) \times10^{-2}$ and $(6.57^{+0.57}_{-1.24})
\times10^{-2}$ for the PSD slope of $\alpha=-1$ and $\alpha=-2$,
respectively. The $\nxsd$ values are thus approximately equal to the
median values of $\nxsb$ for both cases of the PSDs.
Figure~\ref{fig:prob} presents the probability distribution of
$\nxsb$, where the $\nxsd$ value and the $\nxsb$ median are also
plotted, showing that the $\nxsd$ value is very close to the peak
(or the median) of the $\nxsb$ sample for both cases. Our
simulations demonstrate that the estimated $\nxs$ are not
significantly affected by the heavy binning of light curves.
Therefore, the $\nxs$ values of the 300-day averaged ASM light
curves (see Section~\ref{sec:results}) are a good estimator of the
integral of the intrinsic PSD.


\section{Relationship between $\nxs$ and PSD}\label{sec:relation}

\subsection{The assumed PSD shape}

The normalized excess variance, $\nxs$, of a light curve is defined
as (e.g., Zhang et al. 2002)
\begin{equation}
\sigma^{2}_{\rm NXS}=\frac{1}{N \bar{x}^2}
\sum_{i=1}^{N}[(x_i-\bar{x})^2-\sigma_i^2] , \label{eq:nxs}
\end{equation}
\noindent where $N$ is the number of bins in the light curve, $x_i$
and $\sigma_i$ are the count rate and its error of the $i_{\rm th}$
bin, and $\bar{x}$ is the un-weighted arithmetic mean of all $x_i$.
The error on $\nxs$ due to Poisson noise is estimated with
formula~[11] of Vaughan et al. (2003).

A light cure is characterized by its duration, $T$, and binsize,
$\dt$. The PSD of the light curve can be estimated at frequencies
between the minimum frequency, $\numin = 1/T$, and the maximum
frequency, $\numax=1/(2\dt)$ (i.e., the Nyquist frequency).

The AGN hard-state PSD (e.g., O'Neill et al. 2005) is defined as the
doubly-broken power law, with slope $0$ below the low-frequency
break ($\lfb$), slope $-1$ between $\lfb$ and the high-frequency
break ($\fb$), and slope $-2$ above $\fb$,
\begin{equation}
P(\nu)=A(\lfb/\fb )^{-1} \quad (\nu < \lfb) \label{eq:hard1} ,
\end{equation}
\begin{equation}
P(\nu)=A(\nu/\fb )^{-1} \quad (\lfb \leq \nu \leq \fb) ,
\end{equation}
\begin{equation}
P(\nu)=A(\nu/\fb )^{-2} \quad  (\nu > \fb) \label{eq:hard3} .
\end{equation}

Here we define the AGN soft-state PSD as the singly-broken power
law, with slope of $-1$ and $-2$ below and above the break
frequency, $\fb$,
\begin{equation}
P(\nu)=A(\nu/\fb )^{-1} \quad (\nu \leq \fb) \label{eq:psd1} ,
\end{equation}
\begin{equation}
P(\nu)=A(\nu/\fb )^{-2} \quad  (\nu > \fb) \label{eq:psd2} .
\end{equation}
\noindent We mention that the break frequency in the soft-state PSD
model is identical to the high-break frequency in the hard-state PSD
model, both are denominated as $\fb$.

Regarding $\fb$, there are two different assumptions. The first one
assumes that $\fb$ inversely scales with $\mbh$ in the form of
\begin{equation}
\fb = C_{\rm b}/\mbh. \label{eq:cb}
\end{equation}
\noindent By fitting the hard-state PSD predictions (see
Section~\ref{sec:hard}) to the short-timescale variance-mass
relation, O'Neill et al. (2005) obtained $C_{\rm b}=43$ (Hz
$\msun$). Using longer timescale variance-mass relation, Papadakis
(2004) also found similar value for $C_{\rm b}$. The second
assumption for $\fb$ is that $\fb$ depends not only on $\mbh$, but
also on bolometric luminosity, $\lbol$. Based on the results from
PSD analysis of the light curves of ten AGNs and two GBHs, McHardy
et al. (2006) obtained the following relation,
\begin{equation}
\log \fb = -2.1\log\mbh + 0.98\log\lbol + 2.32 \label{eq:tml} ,
\end{equation}
\noindent which shows that $\fb$ (in unit of day$^{-1}$) roughly
inversely scales with the square of $\mbh$ (in unit of $10^6\msun$)
and roughly linearly scales with $\lbol$ (in unit of $10^{44}$erg
s$^{-1}$) from GBHs to AGNs. Note that these two assumptions
(formulae~[\ref{eq:cb}[ and~[\ref{eq:tml}]) are not consistent with
each other, although $\fb$ has stronger dependence on $\mbh$ than on
$\lbol$ in the second assumption.

In the hard-state PSD model, the ratio of $\fb$ to $\lfb$, marked as
$\clfb$, is assumed to be same for all AGNs. Because the
short-timescale variance-mass relation can not constrain the value
of $\clfb$, O'Neill et al. (2005) fixed $\clfb = 20$ in their fits,
assuming that AGNs have similar values of $\clfb$ to those of GBH
hard-state PSDs. Papadakis (2004) also found similar value.

The PSD normalization factor $A$ is the power at $\fb$. The
"universal PSD amplitude", defined as $\psdamp=A\fb$, is assumed to
be the same for all AGNs. Based on the fit to the variance-mass
relation with the hard-state PSD prediction, O'Neill et al. (2005)
obtained $\psdamp=0.024$, similar to the one found by Papadakis
(2004). Nevertheless, constant $\psdamp$ (i.e., same for all
objects) is an assumption only. If $\psdamp$ is not constant, then
formulae~[\ref{eq:soft1}] and~[\ref{eq:hard1}] should not predict a
constant $\nxs$ either.

By definition, $C_{\rm b}$ and $\psdamp$ has the same values in the
two PSD models. Therefore, the values of $\psdamp$ and $C_{\rm b}$,
derived from the fits of the hard-state PSD predictions to the
observed variance-mass relations (O'Neill et al. 2005; Papadakis
2004), also apply to the soft-state PSD model.


\subsection{Model predictions on excess variance}

If the hypothesized PSDs are able to describe the X-ray variability
of AGNs, the observed variance of a light curve is approximately
equal to the predicted variance by integrating the PSD over the
frequency range between $\numin$ and $\numax$,
\begin{equation}
\sigma^2_{\rm NXS} \approx \int_{\numin}^{\numax} P(\nu)d\nu .
\end{equation}
\noindent The predicted $\nxs$ can be analytically expressed in
terms of the parameters of both the light curve and the PSD,
depending on where the break frequencies locate with respect to
$\numin$ and $\numax$.


\subsubsection{The soft-state PSD predictions}\label{sec:soft}

If $\numin \geq \fb$, the predicted variance, integrated the PSD
over slope $-2$ range only, linearly scales with $\fb$,
\begin{equation}
\nxs=\psdamp (T-(2\dt)) \fb . \label{eq:soft2}
\end{equation}

If $\numax \leq \fb$, the predicted variance, integrated the PSD
over slope of $-1$ range only, is independent of $\fb$,
\begin{equation}
\sigma^2_{\rm NXS}=\psdamp (\ln T -\ln(2\dt)) . \label{eq:soft1}
\end{equation}

If $\numin<\fb<\numax$, the predicted variance is
\begin{equation}
\nxs=\psdamp (\ln\fb + \ln T - (2\dt)\fb +1) . \label{eq:soft12}
\end{equation}


\subsubsection{The hard-state PSD predictions}\label{sec:hard}

If $\numin \geq \fb$, the predicted variance, integrated the PSD
over slope $-2$ range only, linearly scales with $\fb$,
\begin{equation}
\nxs=\psdamp (T-(2\dt)) \fb , \label{eq:hard2}
\end{equation}
\noindent which is the same to formula~[\ref{eq:soft2}].

If $\lfb \leq \numin < \numax \leq \fb$, the predicted variance,
integrated the PSD over slope of $-1$ range only, is independent of
$\fb$,
\begin{equation}
\sigma^2_{\rm NXS}=\psdamp (\ln T -\ln(2\dt)) , \label{eq:hard1}
\end{equation}
\noindent which is the same to formula~[\ref{eq:soft1}].

If $\lfb \leq \numin<\fb<\numax$, the predicted variance is
\begin{equation}
\nxs=\psdamp (\ln\fb + \ln T - (2\dt)\fb +1) , \label{eq:hard12}
\end{equation}
\noindent which is the same to formula~[\ref{eq:soft12}].

If $\numin < \lfb < \numax \leq \fb$, the predicted variance is
\begin{equation}
\nxs = \psdamp (1 + \ln\clfb - \ln\fb - \ln(2\dt) - \clfb T^{-1}
\fb^{-1} ) . \label{eq:hard01}
\end{equation}

If $\numin < \lfb < \fb < \numax$, the predicted variance is
\begin{equation}
\nxs = \psdamp (2 + \ln\clfb - \clfb T^{-1} \fb^{-1} - (2\dt)\fb ) .
\label{eq:hard012}
\end{equation}

Finally, if $\numax \leq \lfb$, the predicted variance is
\begin{equation}
\sigma^{2}_{\rm NXS} = \psdamp \clfb ((2\dt)^{-1} - T^{-1}) \fb^{-1}
. \label{eq:hard0}
\end{equation}
\noindent In this case, the variance linearly inversely scales with
$\fb$ when it is measured over $\alpha=0$ part of the PSD only.
Obviously, this is the most strongest signature for the presence of
the second, low break ($\lfb$) at which the PSD flattens from
$\alpha=-1$ to $\alpha=0$.


\subsection{Dependence of excess variance on $\mbh$ and $\lbol$}

If $\fb$ scales with $\mbh$ only, the PSD predictions
(formulae~[\ref{eq:soft2}-\ref{eq:hard0}]) can be further expressed
in terms of $\mbh$ by substituting $\fb$ with $\cb/\mbh$. In such a
way, if $\fb < \numin$, the variance linearly inversely scales with
$\mbh$ in both the PSD models. The hard-state PSD predicts that the
variance linearly scales with $\mbh$ if $\numax < \lfb$, which does
not exist for the soft-state PSD.

On short timescales, even if $\numin < \fb$ (but $\numin > \lfb$),
the two PSD models predict the same variance-mass relation (see
formula~[\ref{eq:soft2}--\ref{eq:hard12}]). Accordingly, the
observed short-term variance-mass relations can not be used to
differentiate the soft-state from the hard-state PSD model. In order
to effectively distinguish the two PSD models, the variance should
be measured with long-term light curve whose length is long enough
to have $\numin < \lfb$, because on this sufficiently long
timescales the soft-state PSD prediction is different from the
hard-state PSD prediction. If the light curves are sampled at
binsize whose corresponding $\numax$ is smaller than $\lfb$, the
hard-state PSD predicts that the variance linearly increase with
$\mbh$ (formula~[\ref{eq:hard0}]), whereas the soft-state PSD still
predicts that the variance is independent of $\mbh$
(formula~[\ref{eq:soft1}]).

If taking into account the scaling of $\fb$ with both $\mbh$ and
$\lbol$, the predicted variance can be further expressed in terms of
$\mbh$ and $\lbol$ by substituting $\fb$ with
formula~[\ref{eq:tml}]. In this case, it is impossible to describe
the variance as the function of mass (or luminosity) separately. For
the correct PSD model, however, the predicted variances would be
equal to the observed variances. Of course, a number of AGNs should
be monitored for sufficiently long time (requiring $\numin < \lfb$)
with regular sampling rate in order to determine whether AGN PSDs
have the second, low-frequency break or not.


\section{Results}\label{sec:results}

Reliable estimation of observed $\nxs$ requires that the light curve
has high signal-to-noise ratio and enough data points. However, the
ASM sensitivity is not sufficient to directly calculate the
variances from the one-day averaged light curves (see discussion of
Section~\ref{sec:obs}). By re-binning the one-day averages over
longer time interval, it is possible to obtain light curves with
good signal-to-noise ratios. Even though the rebinning reduces the
number of data points and conceals short-term variability, it is
still meaningful to acquire long-timescale variability of AGNs. This
is also in accordance with our purpose to estimate long-timescale
variances for differentiating the two PSD models hypothesized for
AGNs. To assure the use of $\chi^2$ statistics and to substantially
increase signal-to-noise ratio, we compromisingly rebin the one-day
averaged light curves with binsize of $\dt=300$~days. The count rate
in each new bin is obtained by weightily averaging all the original
one-day averages in that bin. This procedure of rebinning results in
17 data points in the 300-day averaged light curve for each source.
The values of $\nxs$ calculated with the re-binned light curves are
listed in Table~\ref{tab:agn}.

The rebinned light curves have binsize of $\dt=300$~days and length
of $T=5100$~days. Their variances would be approximately equal to
the predicted variances by integrating the PSD between
$\numin=2.27\times10^{-9}$~Hz and $\numax=1.93\times10^{-8}$~Hz.
More importantly, formula~[\ref{eq:tml}] predicts that the lowest
value of $\fb$ is $\sim 1.22\times10^{-7}$~Hz ($\sim
1.50\times10^{-7}$~Hz if $\fb$ scales with $\mbh$ only) for the
largest mass object 3C~390.3 in our AGN sample. Therefore, the
binsize of $300$~days guarantees that $\numax$ is significantly
lower than $\fb$ for all of our AGNs. As a result, the soft-state
PSD predicts that the observed variances of our AGNs would be the
same and thus independent of $\mbh$. In the hard-state PSD model,
however, the assumption of $\fb/\lfb \sim 20$ implies $\numax <
\lfb$ for AGNs with $\mbh \lesssim 2\times 10^8 \msun$. The observed
variances of these AGNs would linearly increase with mass. The
300-day rebinned light curves are therefore more effective to
distinguish the two AGN PSD models.


\subsection{Relationship between $\nxs$ and $\mbh$} \label{sec:mass}

Figure~\ref{fig:mass} shows the relationship between the estimated
$\nxs$ and $\mbh$. BLS1 and NLS1 do not occupy different regions.
Although the scatter is large, it appears that the variance does not
depend on mass. The Pearson's correlation coefficient between $\log
\nxs$ and $\log \mbh$ is $r=-0.23$, and the null probability is
$p=0.25$, suggesting that the variance may not correlate with
$\mbh$.

The blue solid line in Figure~\ref{fig:mass} plots the variance-mass
relation predicted by the hard-state PSD model, in which we adopt
the best fit parameters ($\psdamp=0.024$, $C_{\rm b}=43$ Hz $\msun$,
and $\clfb=20$) obtained by O'Neill et al. (2005) with the short
timescale variance-mass relation. It is clear that the predicted
relation is not in agreement with the observed one at all. In fact,
the hard-state PSD predicts that the variances linearly increase
with mass for $\mbh \lesssim 10^{8}\msun$. The predicted
variance-mass relation flattens when $\mbh \gtrsim
2\times10^{8}\msun$. This means that AGNs of $\mbh \lesssim
10^{8}\msun$ have $\numax < \lfb$, whereas those of $\mbh \gtrsim
2\times 10^{8}\msun$ have $\numin < \lfb < \numax$.

However, the failure of the hard-state PSD prediction merely shows
that a very specific version of the hard-state PSD model does not
fit the data. This does not prove that the hard-state PSD model,
with another set of parameter values can not fit the data (to some
extent). Among the three parameters, $\cb$, the scaling of $\fb$
with mass, is best determined and mostly impossible to change a lot.
As we will discuss in Section~\ref{sec:scatter}, the value of
$\psdamp$, primarily due to the intrinsic scatter of variance
measurements, may range from 0.008 up to 0.052. In
Figure~\ref{fig:mass}, we plot again the hard-state PSD predictions
by changing the values of $\psdamp$ to 0.008 and 0.052. The
$\psdamp=0.008$ prediction (the bottom blue dashed line) fails
completely. Although the $\psdamp=0.052$ prediction is on top of the
$\psdamp=0.024$ one, it still can not "fit" adequately the data. As
a result, it is rather unlikely that the hard-state PSD predictions
can explain the data by altering the values of $\psdamp$.

In fact, the hard-state PSD predictions, within the $\psdamp$ range
adopted above, are about $1-2$ order of magnitude smaller than the
observed variances for $\mbh \lesssim 10^8 \msun$. This indicates
that, down to frequencies of $\sim 10^{-9}$~Hz, the PSDs of these
AGNs should not show the second low-frequency breaks. In order to
test this assumption, we show in Figure~\ref{fig:mass} (the magenta
solid line) the hard-state PSD prediction by setting $\clfb=2000$,
which explains the data much better than the case of $\clfb=20$.
Thus, our long timescale variance-mass relation does not favor
$\clfb=20$ assumed by O'Neill et al. (2005) for the short timescale
variance-mass relation. However, our data show that the second
breaks of AGN PSDs, if exist, should be close to or beyond the
frequency of $\numin \sim 2.27\times10^{-9}$~Hz. Similarly, the low
frequency end reached by the presently known high quality AGN PSDs
(e.g., McHardy et al. 2004, 2005) have extended through to $\sim
10^{-8}$~Hz, at which the PSDs do not show the second breaks either.
Hence, both the variance and PSD data imply that the ratio of $\fb$
to $\lfb$ should be larger than $\sim 2000$ if the second breaks of
AGN PSDs would exist.

The black dashed line in Figure~\ref{fig:mass} plots the soft-state
PSD prediction by using $C_{\rm b}=43$~Hz~$\msun$ and $\psdamp
=0.024$, same as the ones used for the hard-state PSD prediction.
The soft-state PSD prediction is independent of mass due to $\numax
< \fb$ for all of our AGNs, which is roughly in agreement with the
data. The mass-independent variance predicted by the soft-state PSD
is $5.14\times10^{-2}$, slightly smaller than the average value
($6.64\times10^{-2}$, the red dashed line) of the observed variances
of 27 AGNs (see Section~\ref{sec:scatter}). The red dashed line is
also identical to the soft-state PSD prediction by using
$\psdamp=0.03$, obtained from the averaged variance with
formula~[\ref{eq:soft1}] (see Section~\ref{sec:scatter}), and
$C_{\rm b}=43$~Hz~$\msun$. Consequently, our long-timescale
variance-mass relation seems to favor the soft-state PSD for AGNs.

With the present data, the soft-state PSD prediction is actually
identical to the hard-state PSD prediction with $\clfb \gtrsim
2000$. However, the value of $\clfb \gtrsim 2000$ is remarkably
larger than the currently approved value ($\sim 10-100$) by analogy
with the GBH hard-sate PSDs. This suggests that such large value of
$\clfb$ would not be physical for AGNs. As a result, our long
timescale variance-mass relation is very likely to agree with the
soft-state PSD prediction rather than with the hard-state PSD
predictions.


\subsection{Relationship between $\nxs$ and $\lbol$}\label{sec:bol}

\begin{figure} \epsscale{0.6}
\plotone{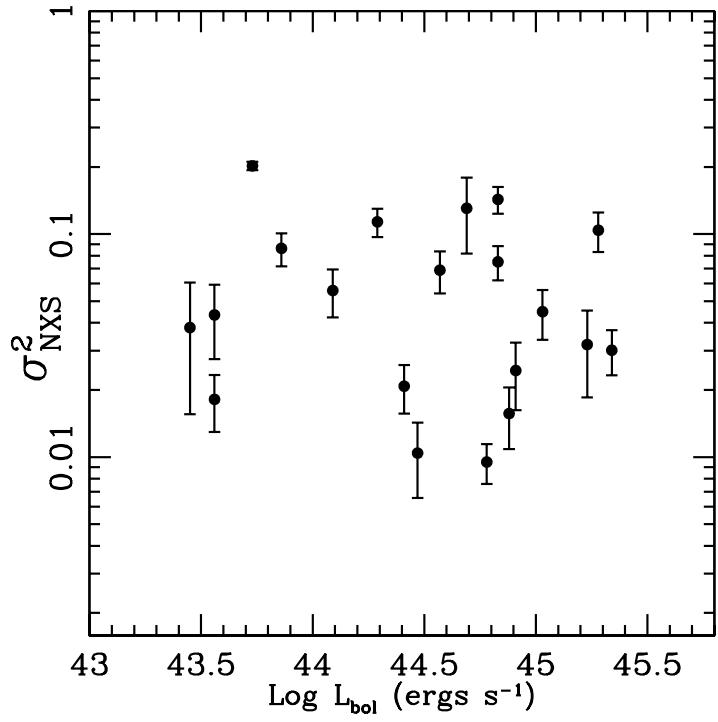} \caption{ \footnotesize The relationship
between the normalized excess variance ($\nxs$) and bolometric
luminosity ($\lbol$). The variance appears to be independent of
luminosity.} \label{fig:lumi}
\end{figure}

On short timescales, the observed variance anti-correlates with
X-ray luminosity of AGNs (e.g., Nandra et al. 1997), whereas it
weakly increases with larger bolometric luminosity (Zhou et al.
2010). Figure~\ref{fig:lumi} plots the relationship between our
long-timescale $\nxs$ and $\lbol$ for 20 AGNs. It appears that the
variance does not depend on $\lbol$. The Pearson's correlation
coefficient between $\log \nxs$ and $\log \lbol$ is $r=-0.09$, and
the null probability is $p=0.70$, indicating that the variance does
not correlate with $\lbol$. This independence of variance on
bolometric luminosity presents another evidence for the soft-state
PSD shape of AGNs. Because all of our AGNs show $\numax < \fb$, the
soft-state PSD deservedly predicts that the variance is independent
of $\lbol$, suggesting that formula~[\ref{eq:tml}] is consistent
with the lack of any correlation between $\nxs$ and $\lbol$.
However, the hard-state PSD with $\clfb=20$ would predict that the
variance decreases with increasing bolometric luminosity.


\subsection{Scatter of $\nxs$ and constant $\psdamp$}\label{sec:scatter}

For the soft-state PSD model, formula~[\ref{eq:soft1}] shows that
the value of $\psdamp$ can be directly estimated from the observed
variance of an object if its $\fb$ is larger than $\numax$. Due to
$\fb > \numax$ for all objects in our sample, their observed
variances, independent of $\mbh$ (and $\lbol$), are expected to be
the same. Therefore, a constant $\psdamp$ (same for all AGNs) can be
obtained with the observed variances of these objects. However,
Table~\ref{tab:agn} (also Figure~\ref{fig:mass}) shows that the
observed variances exhibit quite large scatter by about one order of
magnitude, indicating that the derived values of $\psdamp$ have an
identical degree of scatter from object to object. This is
inconsistent with the assumption that the value of $\psdamp$ is the
same for all AGNs. However, the large scatter of the observed
variances is probably due to the intrinsic scatter (probably very
large) of the variance measurements (e.g., see Section~5 in Vaughan
et al. 2003 and the entries in their Table~1, also see Section~3.2
in O'Neill et al. 2005). Consequently, the scatter of the estimated
variances does not imply the intrinsic scatter of $\psdamp$. The way
to reduce the intrinsic scatter of variance measurements is to
average a number of estimated variances. As a result, in order to
obtain the constant $\psdamp$ and reduce its uncertainty, we average
the observed variances of our 27 AGNs and derive its standard
deviation, which is $6.44\pm4.76 \times 10^{-2}$. With
formula~[\ref{eq:soft1}], we obtain $\psdamp=0.030\pm0.022$. Based
on the short-timescale variance-mass relations, not sensitive to
differentiate the soft-state and hard-state PSD predictions,
Papadakis (2004) and O'Neill et al. (2005) obtained $\psdamp \sim
0.02-0.03$, similar to the one we obtained.

The AGN PSDs obtained with high-quality RXTE and XMM-Newton light
curves indicate that the values of $\psdamp$ also show large scatter
by up to an order of magnitude for different AGNs (e.g., Markowitz
et al. 2003; Done \& Gierli\'nski 2005; Uttley \& McHardy 2005).
Most probably, the uncertainty in the best fit $\psdamp$ does not
indicate a range of intrinsic $\psdamp$, but rather indicates the
true uncertainty with which one can estimate the constant $\psdamp$
with the present PSDs. Nevertheless, one could average the known
values of $\psdamp$ from the best fits to the high-quality PSDs of a
few AGNs as an approximate value of the constant $\psdamp$. The
averaged value of $\psdamp$ ($\sim 0.02-0.03$) is also similar to
the one we obtained from the average of the observed variances of a
number of AGNs under the assumption of the soft-state PSD shape.


\subsection{The $\nxs - \mbh - \lbol$ plane}

Formula~[\ref{eq:tml}] shows that $\fb$ depends on both $\mbh$ and
$\lbol$. It is therefore inadequate to study the $\nxs - \mbh$ and
$\nxs - \lbol$ relation, respectively. If taking into account the
scaling of $\fb$ with $\mbh$ and $\lbol$ synchronously, we have to
compare the observed variances with the ones predicted by the PSD
models, i.e., a projection of the $\nxs - \mbh - \lbol$ plane. The
predictions are calculated with the soft-state and hard-state PSD,
respectively, with the values of $\fb$ calculated with
formula~[\ref{eq:tml}] and $\psdamp=0.03$ estimated in
Section~\ref{sec:scatter}. For the hard-state PSD predictions,
$\clfb$ is still assumed as $20$.

\begin{figure} \epsscale{0.6}
\plotone{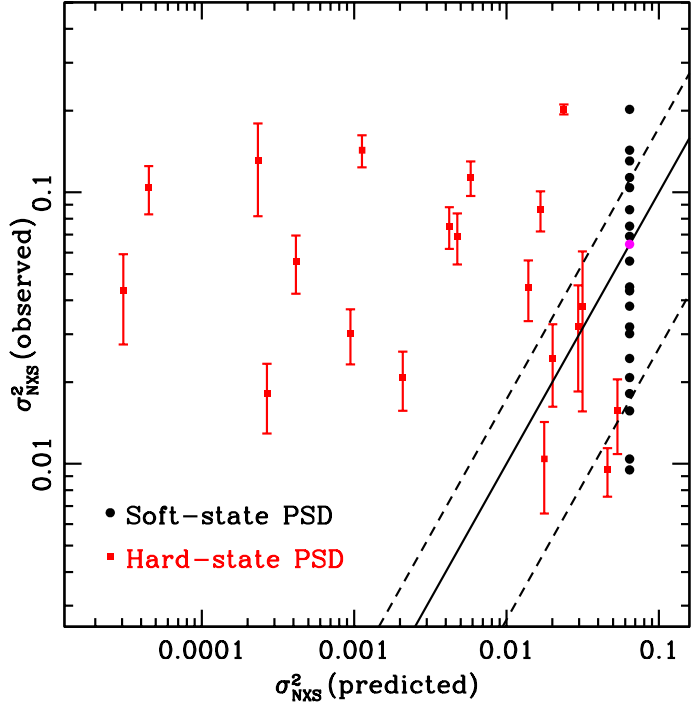} \caption{ \footnotesize The observed $\nxs$ is
plotted against the predicted $\nxs$ for the soft-state (black solid
circles) and hard-state (red solid squares) PSD models,
respectively. For the soft-state and hard-state PSD predictions, we
adopt $\psdamp=0.03$, and the PSD break frequency ($\fb$) scales
with black hole mass and bolometric luminosity via
formula~[\ref{eq:tml}]. The ratio of $\fb$ to $\lfb$ ($\clfb$) is
assumed to be $20$ for the hard-state PSD model. The black solid
line is not a best fit line to the relation between the observed and
predicted variances, whereas it is the position on which an object
will lie if the predicted variance is equal to the observed one. The
two black dashed lines indicate $1 \sigma$ deviation from the solid
line, representing the positions to which the solid line will shift
if the predicted variance is calculated with $\psdamp=0.052$ (the
upper black dashed line) and $\psdamp=0.008$ (the lower black dashed
line), respectively. The predicted variances by the soft-state PSD
are the same for all of the objects (the magenta solid circle
indicates the averaged value of the observed variances). The error
bars on the observed variances are not shown for the soft-state PSD
model, but they are identical to the ones for the hard-state PSD
model since the observed variances are the same for the two PSD
models. Most of AGNs locate far away from the three lines for the
hard-state PSD predictions, whereas they lie close to the lines for
the soft-state PSD predictions. } \label{fig:obspred}
\end{figure}

Figure~\ref{fig:obspred} plots the observed $\nxs$ against the
predicted $\nxs$ (the black solid circles for the soft-state PSD
predictions and the red solid squares for the hard-state PSD ones)
for 20 AGNs. We specially emphasize that the black solid line in
Figure~\ref{fig:obspred} is not a best fit line to the relation
between the observed and predicted variances (for the two PSD
models, respectively). It is the line on which an object will lie
exactly if the predicted variance is equal to the observed one. In
the same way, the two black dashed lines indicate $1 \sigma$
deviations from the black solid line, showing the positions to which
the black solid line will shift if the predicted variances are
calculated with $\psdamp=0.052$ (the upper black dashed line) and
$\psdamp=0.008$ (the lower black dashed line), respectively. In
other words, the upper and bottom black dashed lines have the
exactly same meanings as the black solid line if the (soft-state or
hard-state) PSD predictions are calculated with $\psdamp=0.052$ and
$\psdamp=0.008$, respectively. The two values of $\psdamp$ are the
upper and lower $1\sigma$ deviations from $\psdamp=0.03$ obtained in
Section~\ref{sec:scatter}.

Due to $\fb > \numax$, the predicted variances by the soft-state PSD
are the same for all of our AGNs (the magenta solid circle indicates
the averaged value of the observed variances). As a result, the
relation between the observed and predicted variances is
perpendicular to the axis of the predicted variance. Most of the
objects lie within the $1\sigma$ range of the black solid line. For
the hard-state PSD, however, most of AGNs locate far away from the
black solid and dashed lines. Therefore, Figure~\ref{fig:obspred}
suggests that the soft-state PSD predictions are much more
consistent with the data than the hard-state PSD predictions, but
the latter is only for the specific parameters of O'Neill et al.
(2005) as already discussed in Section~\ref{sec:mass}. Moreover, for
the hard-state PSD predictions, the majority of the most massive
objects lie close to the black solid line, indicating that their
values of $\numin$ are slightly smaller than the respective values
of $\lfb$. As a result, these massive AGNs, due to longer
timescales, are not good indictors for differentiating the two PSD
models.


\section{Discussion and conclusions}\label{sec:disc}

With the ASM data accumulated over 14 years, we estimate the
normalized excess variances of 27 AGNs in the X-ray band on the
longest timescale ($\dt=300$~days and $T=5100$~days) so far.
Although the observed variances show quite large scatter, they
appear to not depend on black hole mass and bolometric luminosity of
AGNs. This phenomenology has already been noticed by Markowitz \&
Edelson (2004) with PCA light curves of $\dt=34.4$~days and
$T=1296$~days. The short-timescale variance-mass relation has been
able to constrain the PSD amplitude and the scaling law of the PSD
high-break frequency with black hole mass (Papadakis 2004; O'Neill
et al. 2005), but it can not be used to effectively distinguish the
hard-state and soft-state PSD models hypothesized for AGNs (as they
have the same predictions on short timescales) in general.
Relatively, the long timescale variance-mass relation is more valid
to determine whether the AGN PSDs have the second low-frequency
breaks or not.

With the best-fit parameters obtained by O'Neill et al. (2005), the
soft-state PSD prediction is much more consistent with our long
timescale variance-mass relation than the hard-state PSD prediction.
On the longest timescale so far, the hard-state PSD predicts that
the variance increases with mass, whereas the soft-state PSD gives
the mass-independent variance. In fact, the variances predicted by
the hard-state PSD model are remarkably smaller than the observed
variances for most of our AGNs. Furthermore, by taking into account
the scaling factor of the PSD high-break frequency with black hole
mass and bolometric luminosity together, the soft-state PSD
predictions are more likely to agree with the observed variances
than the hard-state PSD predictions. Therefore, the long timescale
variances appear to favor the soft-state rather than the hard-state
PSD shape for AGNs. It is reasonable to assume that it is the
variances predicted by the best-fit PSD parameters of O'Neill et al.
(2005), not by the hard-state PSD in general, that are not
consistent with the data. However, we have demonstrated that changes
of PSD amplitude within the most possible range do not improve the
"fit" to the data. Rather, the hard-state PSD model could probably
agree with the data, but it would require a range of PSD amplitude
values that have not been observed so far in the cases when a proper
fitting has been done to the PSDs of a few sources. Despite the
hard-state PSD predictions with $\fb / \lfb \gtrsim 2000$ can
explain the data as good as the soft-state PSD predictions, such
large value of $\fb / \lfb$ is very probably nonexistent by scaling
from GBHs to AGNs. Moreover, the hard-state PSD with $\fb / \lfb
\gtrsim 2000$ is actually identical to the soft-state PSD for the
present data.

\begin{figure} \epsscale{0.6}
\plotone{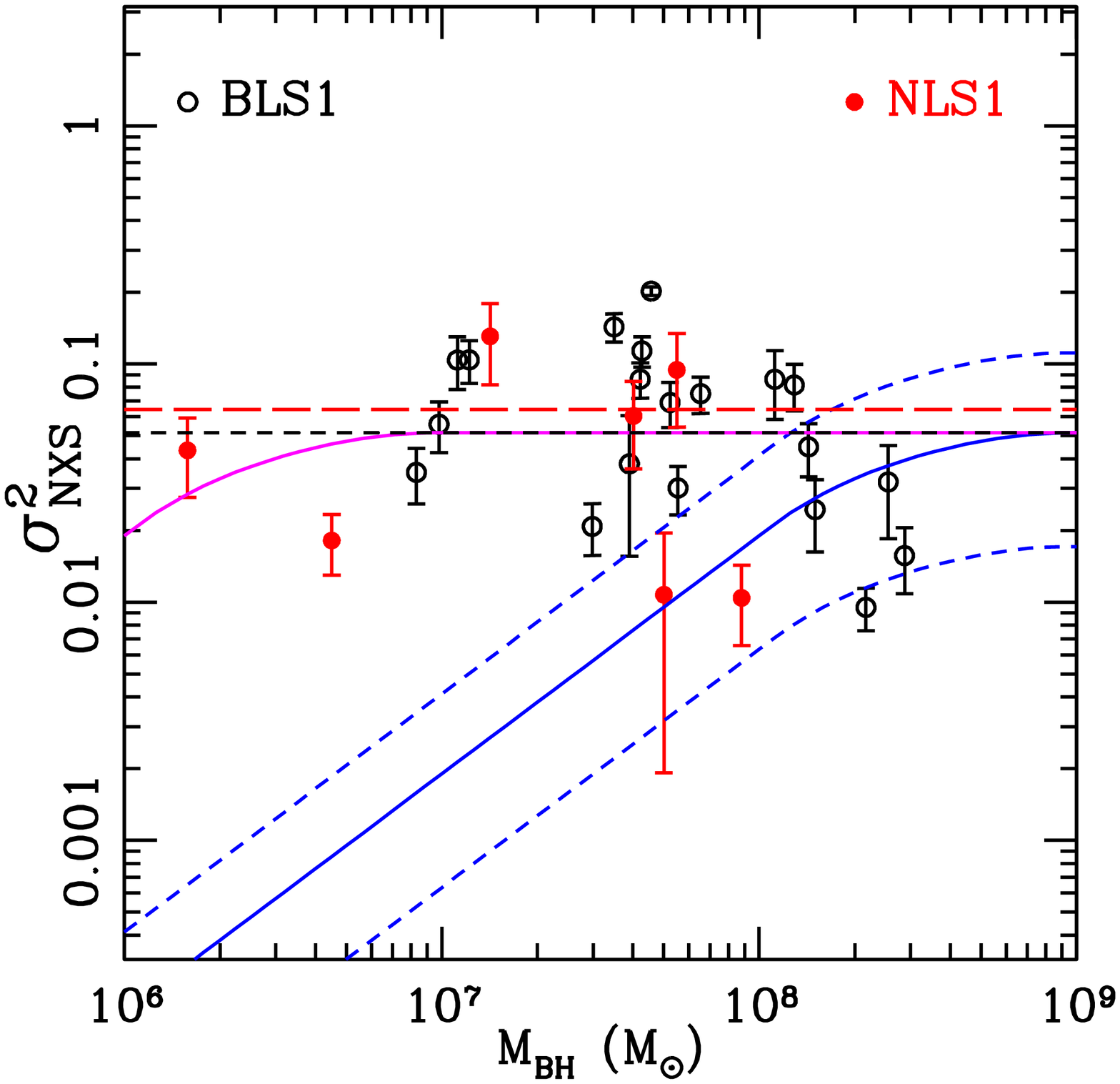} \caption{ \footnotesize  The relationship
between the normalized excess variance ($\nxs$) and black hole mass
($\mbh$). It appears that the variance does not depend on the mass.
For the PSD predictions shown in this figure, $\fb$ scales with
$\mbh$ only, and $C_{\rm b}=43$~Hz~$\msun$ is assumed.
The blue solid line is the hard-state PSD prediction with
$\psdamp=0.024$ and $\clfb=20$. The upper and bottom blue dashed
lines correspond to the hard-state PSD predictions by changing the
value of $\psdamp$ to $0.052$ and $0.008$ (keeping $\clfb=20$),
respectively. The magenta solid line shows the hard-state PSD
prediction by setting $\clfb=2000$ (holding $\psdamp=0.024$).
The black dashed line is the soft-state PSD prediction with
$\psdamp=0.024$. The red dashed line shows the averaged value of the
observed variances of 27 AGNs, which is the same to the soft-state
PSD prediction with $\psdamp=0.03$ obtained from the average
variance with formula~[\ref{eq:soft1}]. It can be seen that the
observed variance-mass relation is much more consistent with the
soft-state PSD predictions (independent of mass) than the hard-state
PSD predictions. The hard-state PSD prediction also "fits" the data
only by increasing $\clfb$ by two order of magnitude. }
\label{fig:mass}
\end{figure}

Up to now, the best fits show that the high-quality PSDs of a few
AGNs have one high-frequency break only except for Ark~564 who
exhibits the second low-frequency break. Thereby, our variance
analysis of a larger sample of AGNs agrees with the direct PSD
analysis of a smaller sample of AGNs, both implying the soft-state
PSD shape for AGNs. For example, the PSDs of NGC~4051 and
MCG-6-30-15, having the smallest mass among our sample, show one
break only down to frequency of $\sim 10^{-8}$~Hz at quite high
confidence level (McHardy et al. 2004; 2005). At the same time,
Figure~\ref{fig:mass} shows that the estimated variances of the two
objects, roughly consistent with the soft-state PSD predictions, are
remarkably larger than the variances predicted by the hard-state PSD
model. Compared to the PSD analysis, however, our variance analysis
lowers the frequencies by more than one order of magnitude.
Therefore, our results further strengthen the standpoint, first
suggested by the PSD analysis, that AGN PSDs might have one break
only.

By analogy with the PSDs of GBHs in the high/soft state, both the
variance and PSD analyzes suggest that Seyfert-type AGNs are in the
high accretion state. Therefore, AGNs are thought to be scaled-up
GBHs in the high/soft state, characterized by the PSD break
frequency scaling with black hole mass and bolometric luminosity (or
Eddington accretion rate) from GBHs to AGNs (McHardy et al. 2006).
Under these assumptions, the break frequencies of all AGNs in our
sample are substantially smaller than the maximum frequency to which
the 300-day binned light curves correspond, indicating that the
estimated variances are the same for our AGNs. This motivate us to
average the observed variances, from which we obtain a constant PSD
amplitude of $\sim 0.030\pm0.022$ for AGNs, roughly consistent with
those obtained from the direct PSD analysis and the short-timescale
variance analysis.

There are a number of factors which may cause the large scatter of
the observed variances. One main effect may come from the low
quality ASM light curves for AGNs (see Section~\ref{sec:obs}). From
the point view of PSD itself, the slope above and below the break
frequency are not exactly equal to $-2$ and $-1$ for most of AGNs,
as already seen from the known AGN PSDs (e.g., Uttley, McHardy \&
Papadakis 2002). At the same time, the AGN PSDs are better described
by a bending power law (e.g., Uttley \& McHardy 2005) rather than by
an abruptly broken power law assumed here for simplicity. It is
worth noting that, if the soft-state PSD model is applicable to
AGNs, the observed variances should be the same. However, because
the variance measurements have the known intrinsic scatter (Vaughan
et al. 2003; O'Neill et al. 2005), which could be very large, the
large scatter of the estimated excess variances is almost certainly
due to the intrinsic scatter of the $\nxs$ values themselves. In
order to decrease the intrinsic scatter of variance measurements, it
is necessary to perform high quality observations on longer
timescale (and/or multiple observations) for AGNs. Averaging the
multiple variances of an object (from segments of long-timescale or
multiple observations) can reduce the intrinsic uncertainty of
variance measurements. Due to shorter timescales, the AGNs with
smaller or intermediate black hole mass are more effective to
differentiate the two PSD models.

In conclusion, the longest timescale X-ray variability suggests that
AGN PSDs have one break only, and further indicates that AGNs are in
the high accretion state.


\acknowledgments

I thank the anonymous referee for the constructive suggestions and
comments that significantly improved the paper. The data used in
this paper are based on quick-look results provided by the ASM/RXTE
team. This research has made use of data obtained through the High
Energy Astrophysics Science Archive Research Center Online Service,
provided by the NASA/Goddard Space Flight Center. This work is
supported by the National Natural Science Foundation of China
(Project 10878011 and 10733010) and by the National Basic Research
Program of China -- 973 Program 2009CB824800.


\clearpage

\begin{deluxetable}{lcccccccc}
\tabletypesize{\scriptsize} \tablecolumns{4} \tabcolsep 3pt
\tablewidth{0pc} \tablecaption{The AGN sample and relevant
parameters} \tablehead{
\colhead{Source} &\colhead{ASM} &\colhead{$\sigma^2_{\rm NXS}$}
&\colhead{$M_{\rm BH}$ } &\colhead{ref.,\tablenotemark{a}}
&\colhead{$T_{\rm b}$ } &\colhead{ref.\tablenotemark{b}}
&\colhead{$L_{\rm bol}$ } &\colhead{ref.\tablenotemark{c} } \\
\colhead{Name} &\colhead{rate} &\colhead{($\times 10^{-2}$)}
&\colhead{($10^7 M_\odot$)} &\colhead{meth.} &(day) & &(${\rm ergs~
s^{-1}}$) & }

\startdata
\multicolumn{9}{c}{Broad line objects}\\
3C 120       &0.22 &3.01$\pm$0.69 &5.55 &1,r  &...   &... &45.34 &9  \\
3C 390.3     &0.16 &1.57$\pm$0.48 &28.7 &1,r  &...   &... &44.88 &9  \\
Ark 120      &0.17 &2.44$\pm$0.82 &15.0 &1,r  &...   &... &44.91 &9  \\
Fairall 9    &0.13 &3.20$\pm$1.35 &25.5 &1,r  &28.9  &14  &45.23 &9  \\
IC 4329a     &0.51 &0.95$\pm$0.19 &21.7 &2,d  &4.6   &2   &44.78 &9  \\
MR 2251-178  &0.19 &3.51$\pm$0.92 &0.83 &3,o  &...   &... &...   &...  \\
Mrk 79       &0.14 &6.89$\pm$1.47 &5.24 &1,r  &...   &... &44.57 &9  \\
Mrk 279      &0.12 &14.3$\pm$1.95 &3.49 &1,r  &...   &... &...   &...  \\
Mrk 290      &0.09 &10.4$\pm$2.59 &1.12 &4,o  &...   &... &...   &...  \\
Mrk 509      &0.19 &4.49$\pm$1.13 &14.3 &1,r  &...   &... &45.03 &9  \\
NGC 526a     &0.13 &8.16$\pm$1.81 &12.9 &3,o  &...   &... &...   &...  \\
NGC 985      &0.10 &8.62$\pm$2.78 &11.2 &4,o  &...   &... &...   &...  \\
NGC 3227     &0.18 &8.63$\pm$1.45 &4.22 &1,r  &0.59  &14  &43.86 &9  \\
NGC 3516     &0.13 &11.3$\pm$1.65 &4.27 &1,r  &5.8   &14  &44.29 &9  \\
NGC 3783     &0.27 &2.08$\pm$0.51 &2.98 &1,r  &2.9   &14  &44.41 &9  \\
NGC 4151     &0.42 &20.2$\pm$0.86 &4.57 &5,r  &9.2   &14  &43.73 &9  \\
NGC 4258     &0.09 &3.81$\pm$2.25 &3.90 &6,m  &513   &14  &43.45 &9  \\
NGC 4593     &0.16 &5.58$\pm$1.35 &0.98 &7,r  &...   &... &44.09 &9  \\
NGC 5548     &0.19 &7.51$\pm$1.31 &6.54 &8,r  &18.3  &14  &44.83 &9  \\
NGC 7469     &0.14 &10.4$\pm$2.11 &1.22 &1,r  &...   &... &45.28 &9  \\
\multicolumn{9}{c}{Narrow line objects}\\
IC 5063      &0.08 &9.44$\pm$4.02 &5.50 &9,d  &...   &... &44.53 &9  \\
MCG-6-30-15  &0.27 &1.81$\pm$0.52 &0.45 &10,o &0.15  &14  &43.56 &14  \\
Mrk 335      &0.07 &13.1$\pm$4.88 &1.42 &1,r  &0.068 &15  &44.69 &9  \\
Mrk 478      &0.09 &6.05$\pm$2.41 &4.02 &11,o &...   &... &...   &...  \\
NGC 4051     &0.12 &4.34$\pm$1.58 &0.16 &12,r &0.019 &14  &43.56 &9  \\
NGC 5506     &0.29 &1.04$\pm$0.39 &8.80 &11,d &0.89  &14  &44.47 &14  \\
PKS 0558-504 &0.11 &1.07$\pm$0.88 &4.50 &13,o &...   &... &...   &...  \\

\enddata
\tablenotetext{a}{References for $\mbh$: 1. Peterson \et (2004); 2.
Markowitz (2009); 3. Morales \& Fabian (2002); 4. Bian \& Zhao
(2003); 5. Bentz \et (2006); 6. Herrnstein \et (1998); 7. Denney \et
(2006); 8. Bentz \et (2007); 9. Woo \& Urry (2002); 10. McHardy \et
(2005); 11. Papadakis (2004); 12. Denney \et (2009); 13. Wang et al.
(2001); 14. Uttley \& McHardy (2005); 15. Ar\'evalo et al. (2008).
The letter in this column indicates the method used to estimate
$\mbh$: r -- reverberation mapping, d -- stellar velocity
dispersion, m -- maser, and o -- other methods. }
\tablenotetext{b} {References for $\tb$: 14. Uttley \& McHardy
(2005); 15. Ar\'evalo et al. (2008).}
\tablenotetext{c} {References for $\lbol$: 9. Woo \& Urry (2002);
14. Uttley \& McHardy (2005).}

\label{tab:agn}
\end{deluxetable}


\end{document}